\documentclass[preprint]{elsarticle}
\usepackage[utf8]{inputenc}
\usepackage[T1]{fontenc}
\usepackage[british,UKenglish,USenglish,english,american]{babel}

\usepackage{balance}
\usepackage{multirow}
\usepackage{numprint}
\usepackage{verbatim}
\usepackage{paralist}
\usepackage{graphicx}
\usepackage[bookmarks=false]{hyperref}
\usepackage{numprint}

\usepackage{framed}
\usepackage[most]{tcolorbox}
\usepackage[framemethod=default]{mdframed}
\usepackage{lineno}

\newcommand{\xpmf}{{cross-platform mobile app development frameworks}}

\newcommand{\xfdb}{XamForumDB }

\newcommand{\xfdbse}{XamForumDB}
\newcommand{\qa}{Q\&A }
\newcommand{\xf}{Xamarin Forum }
\newcommand{\xfse}{Xamarin Forum}
\newcommand{\rosen}{Rosen and Shihab }
\newcommand{\rosense}{Rosen and Shihab}
\newcommand{\so}{Stack Overflow }
\newcommand{\sose}{Stack Overflow}
\newcommand{\ent}[1]{{\it{#1}}}
\newmdenv[linecolor=black,backgroundcolor=gray!30]{myframe}

\npdecimalsign{.}
\npthousandsep{,}
\newcommand{\qt}[1]{{\emph{``#1"}}}

\newcommand{\questionamount}{What is the number of  a) Xamarin-related questions, b) answers and accepted, c) views on \so and \xfse?}
\newcommand{\rqamount}{RQ 1: \questionamount}

\newcommand{\rqtopics}{RQ 2: What are the main topics discussed about Xamarin in \qa sites?}
\newcommand{\rqnativet}{RQ 3: How many main topics from  \so and \xf are also topics related to general mobile development discovered by \rosense?}
\newcommand{\rqquestions}{RQ 4: What are the most relevant questions from Xamarin-related topics?}

\begin{document}

\title{Two Datasets of Questions and Answers for Studying  the Development of Cross-platform Mobile Applications using Xamarin Framework}

\begin{frontmatter}
\author{Matias Martinez}
\ead{matias.martinez@univ-valenciennes.fr}
\cortext[cor1]{Corresponding author}
\address{University of Valenciennes, France}

\begin{abstract}

{\bf Context:}
A \emph{cross-platform mobile application} is an application that runs on multiple mobile platforms (Android,  iOS).
Several frameworks, have been proposed to simplify the development of cross-platform mobile applications and, therefore, to reduce development and maintenance costs.
Between them, the \emph{cross-compiler} mobile development frameworks, such as Xamarin from Microsoft, transform the application's code written in intermediate (aka non-native) language to native code for each desired platform.
However, to our best knowledge, there is no much research about the advantages and disadvantages of the use of those frameworks during the development and maintenance phases of mobile applications.

{\bf Objective:}
The objective of this paper is twofold.
Firstly, to present two datasets of questions and answers (Q\&A)   related to the development of mobile applications using Xamarin.
Secondly, to show their usefulness, we present a replication study for discovering the main discussion topics of Xamarin development.

{\bf Method:}
We created the two datasets by mining two Q\&A sites: \xf and \sose.
Then, for discovering the main topics of  the questions from both datasets, we replicated a study that applies Latent Dirichlet Allocation (LDA).
Finally, we compared the discovered topics with those topics about general mobile development reported by a previous study.

{\bf Results:}
Our datasets have 85,908 questions mined from the  \xf  and 44,434 from   \sose.
Between the main topics discovered from those questions, we found that some of them are exclusively related to  Xamarin and Microsoft  technologies such as the  design pattern `MVVM'.

{\bf Conclusions:}
Both datasets with Xamarin-related Q\&A can be used by the research community for understanding the main concerns about developing cross-platform mobile applications using Xamarin framework. In this paper, we used it for replicating a studying about topic discovering.

\end{abstract}


\end{frontmatter}

\section{Introduction}

Nowadays, there are billions of smartphone devices around the world.
Smartphones are mobile devices that run software applications (apps) such as games, social network and banking apps. 
A \emph{native mobile application} is an app built to run in a particular mobile platform.  
Currently, there are two platforms that dominate the smartphone market: Android (from Google) and iOS (from Apple),  with the 99.7\% of the market share as of the first quarter of 2017 \cite{mobileshare}.

A \emph{cross-platform mobile application} is an application that targets more than one mobile platform. 
To cover a large number of users and, thus, to increase the impact on the market and revenues, companies and developers aim at releasing their mobile apps to both Android and iOS platforms. 
A traditional approach for developing this kind of apps is to build, for each platform, a \emph{native} application using a particular programming language (e.g., Java for Android, Objective-C or Swift for iOS), SDK (Software Development Kit) and IDE (e.g., Android Studio, XCode for iOS).
Unfortunately, 
this approach increases the cost of development and maintenance. 
For example, a company needs developers with different competence for developing an app for two platforms, resulting in two native apps. 
Moreover, as studied by previous works, those native apps could have different quality. For example, Hu et al. \cite{hu2016crossconsistency} found that 68\% of the studied cross-platform apps have different start ranking across the App Store and Google Play stores. Furthermore,  Ali et al.  \cite{Ali2017SAD} analyzed \numprint{80000} cross-platforms apps from those app stores and  found that the Android version of app-pairs receives higher user-perceived ratings compared to the iOS version.

During last years, researches (e.g., \cite{perchat2014common}) and industry companies (e.g,  Microsoft, Facebook Inc.) have both focused on proposing development frameworks with the goal of making easier the development and maintenance of cross-platforms mobile apps.
Earliest frameworks focused on producing \emph{hybrid mobile applications}: apps built coding both non-native (e.g., HTML for Phonegap/Cordova\footnote{\url{https://cordova.apache.org}}) and native code. 
The non-native code is shared across all the platforms' implementations, whereas the native is written for a particular platform.
Nonetheless, beyond the good results of some of them 
for developing simple apps (\cite{Joorabchi2013Challenges,heitkotter2012evaluating}), companies such as Facebook found the resulting applications do not have the same user experience than \emph{purely} native applications \cite{Martinez:2017:TQI}.

Another family of frameworks for developing cross-platform apps is the \emph{cross-compiler mobile development framework}.
They generates native code (for a particular platform) from an application written in a non-native language.
Xamarin\footnote{\url{https://www.xamarin.com}} and React-Native\footnote{\url{https://facebook.github.io/react-native}} are two of them, which consider C\# and JavaScript as non-native languages, respectively.
Using cross-compiler frameworks for developing cross-platform mobile applications, developers can reduce both development and maintenance costs by sharing code between the different platform implementations and, at the same time, they obtain pure native mobile applications.

However, beyond the advantages of sharing code, we wonder whether developing cross-platforms apps using cross-compiler frameworks has any drawback with respect to the native development during the life-cycle of a mobile application. 
To our knowledge, as also mentioned by \cite{Nagappan2016:TrendsMobile}, no previous work has studied that yet. 
Our long term goal is to study the differences  between  the development process that uses cross-compiler framework and the process that uses traditional development toolkits.

\begin{myframe}
To encourage the study of  cross-platforms mobile application, this paper presents two datasets of  questions and answers (Q\&A)  related to one cross-platform development framework:  Xamarin from Microsoft.
\end{myframe}

The two datasets are conformed by  questions  and answers extracted from \so and \xfse, respectively. 
The latter is a \qa site exclusively dedicated to Xamarin technology.\footnote{\url{https://forums.xamarin.com/}} 
The main reasons that motivate us to study Xamarin are:
\begin{inparaenum}[\it a)]
\item availability of documentation and guidelines\footnote{\url{https://developer.xamarin.com/guides/}};
\item 5+ books edited during last years (e.g., \cite{hermes2015xamarin,snider2016mastering,peppers2015xamarin});
\item development toolkits available\footnote{\url{https://docs.microsoft.com/visualstudio/}};
\item availability of testing environment for cross-platform apps called Xamarin Test Cloud.\footnote{\url{https://www.xamarin.com/test-cloud}}, and
\item Xamarin is one of the top 10 most popular development frameworks and libraries  according to the \so Survey 2018.\footnote{\url{https://insights.stackoverflow.com/survey/2018/\#technology-frameworks-libraries-and-tools}}

\end{inparaenum}

Previous works have analyzed mobile-related \qa from \so with different purposes, for instance, for discovering main topics related to mobile development (\cite{Linares-Vasquez2013EAM,Rosen2016MDA,Beyer2014}).  
To show the utility of the two  Xamarin-related \qa dataset, in this paper we replicate one of those studies,  i.e., \rosen   \cite{Rosen2016MDA}, focusing exclusively on the Xamarin technology, instead of focusing on general mobile development as done by \rosense.   
Our study discovers the main discussion topics present in the questions related to Xamarin by applying Latent Dirichlet Allocation (LDA).
Finally, to know about the particularities of cross-platform apps developed using Xamarin,  we compare the  topics we discovered with those  topics related to general mobile development discovered by \rosense.

\begin{comment}
Previous works have analyzed \qa from \so for discovering main topics related to native mobile development (\cite{Linares-Vasquez2013EAM,Rosen2016MDA,Beyer2014}).

\end{comment}

Our research is guided by the following research questions:
\begin{enumerate}
\item[] \rqamount
\item [] \rqtopics
\item []{\rqnativet}
\item []{\rqquestions}
\end{enumerate}

\begin{comment}
Our main results show that \xf has a larger number of questions than \sose, however, the latter has more answers per question and a larger number of questions with at least one accepted answer.  
Moreover, \xf and \so share most of the main topics, which mainly discuss about user interface (UI), formatting, design and navigation. 
Finally, we found topics related to Xamarin technology such as the design pattern `MVVM' that were not found in the topics from native mobile development from \rosense.
\end{comment}

The contributions of this paper are:
\begin{enumerate}
\item A dataset of \qa related to Xamarin technology filtered from \sose.

\item A dataset of \qa extracted from Xamarin forum.

\item An analysis of the two datasets (e.g., number of questions, answers, views).

\item Discussion topics  discovered from questions related to Xamarin technology.

\item An study about the relation between topics from \so and \xf and those topics discussed in questions related to development of native mobile applications from \cite{Rosen2016MDA}.

\item The study of the relevant questions from 3 Xamarin-related topics.
\end{enumerate}

The paper is organized as follows.
Section \ref{sec:relatedwork} discusses the related work.
Section \ref{sec:datasets} presents two datasets of Q\&As extracted from \so and \xfse. 
Section \ref{sec:analyzingqa} analyzes both datasets in terms of number questions, answers accepted, and views.
Section \ref{sec:replication}  replicates the study of \rosen  \cite{Rosen2016MDA}  for discovering topics from  \so and \xfse.
Section \ref{sec:discussion} presents the discussion. 
Section \ref{sec:conclusion} concludes the paper.

All data discussed in this paper, including the two \qa datasets, is publicly available in our web appendix: 
\url{http://anonymous.4open.science/repository/3ac646ee-03f9-40d0-a7fc-d0a6124af979/}

\begin{comment}
Xamarin has an official online forum\footnote{https://forums.xamarin.com/}, which has multiple purpose: 
\begin{inparaenum}[a)]
\item to be a platform for communicating announcements (events, releases, jobs related with the platform) 
and 
\item to be a Q\&A site, where Xamarin users (such as cross-platforms developers) post discussions, questions and share answers. 
Also, the developers of the Xamarin platforms (i.e., members of the company that develop it) participate on the forum.
\end{inparaenum}

We consider that \xfdb gives to the software engineering research community the possibility to study the development and maintenance phases of cross-platform mobile applications, and to propose new approaches for improving the quality of mobile applications.

\end{comment}

\section{Related Works}
\label{sec:relatedwork}

During the recent years several works have studied  Q\&A from the site \so \cite{MSRChallenge2013, Bazelli2013PTS,Pinto2014MQS,Pinto2015SMP}. 
Some of them focused on questions about Android platform \cite{Linares-Vasquez:2014:ACT,Wang:2013:DAU,Stevens:2013:APU,Beyer2014,Beyer2016GAT,Abdalkareem2017Reuse, Treude2017CF,Syer2015}.
For example, 
Linares-Vasquez  et al. \cite{Linares-Vasquez:2014:ACT} studied questions and activities in \so when changes on Android APIs occur, finding that deleting public methods from APIs is a trigger for questions that are discussed more.
Wang   et al.  \cite{Wang:2013:DAU} analyzed posts from \so  related to iOS and Android  APIs to find API usage obstacles, and then they applied a topic modeling technique to discover  several repetitive scenarios in which API usage obstacles occur. 
Stevens  et al. \cite{Stevens:2013:APU}  studied  questions about Android permission use on \sose.
Other works have focusing on mobile-related tags from \sose: 
Beyer et al.  \cite{Beyer2014} investigated 450 Android-related posts from \so to get insights into the issues of Android app development. 
To our knowledge, no work has studied \so questions and problematic related to neither cross-complied frameworks nor Xamarin technology.

In this paper, we analyze \qa related to Xamarin technology, and then we apply applied a topic modeling techniques called Latent Dirichlet Allocation (LDA) to obtain the main topics from those questions.
As reported by  Chen et al. \cite{Chen2016survey} and Sun et al. \cite{Sun2016Survey} several works have  already applied topic modeling techniques such as LDA.
For instance,  Barua et al. \cite{Barua2014DTA} analyzed \so data to automatically discover the main topics present in developer discussions. 
Yang et al. \cite{Yang2016}  focused on discovering topics related to security by analyzing  security-related posts. 
Bajaj et al. \cite{Bajaj:2014}  discovered  topics  from discussions about web developers and focused on how prevalent are web-related topics in discussions related to mobile web development.
Other works have focused on topic modeling for native mobile technologies. 
For example, 
Linares-Vasquez et al. \cite{Linares-Vasquez2013EAM} used LDA to extract hot-topics from mobile-development related questions.
Their findings suggest that most of the questions include topics related to general questions and compatibility issues, whereas the most specific topics are present in a reduced set of questions.
\rosen \cite{Rosen2016MDA} applied LDA on mobile-related \so posts to determine what mobile developers are asking. 
Across all native platforms studied, they found that questions related to app distribution, user interface, and input are among the most popular. 
Pinto et al. \cite{Pinto2016Swift} have analyzed questions about Swift, programming language (successor of Objective-C) for building native iSO mobile apps. They applied LDA to find common problems faced by Swift developers, finding that the language is easy to understand and adopt, but there are many questions about problems in the toolkit (IDE, SDK).
In our work, we focus on topics related to Xamarin development framework.

Other works have study both  \so and other sources of information.
For example, Wang et al. \cite{Wang2017Linking} studied the mutual knowledge sharing between Android Issue Tracker and \sose. Their goal is to bridge the two communities by linking related issues to posts automatically. 
Ye et al. \cite{Ye2017DK} studied how users share URLs in \so for understanding how knowledge diffusion process takes place on that site. 
They found that the 31\% of the shared URLs on \so is to reference information that can help to solve a complex problem. 
Other works focus on analyzing developer forums.
For example, Venkatesh et al.  \cite{Venkatesh2016} mined both developer forums and \so to find the common challenges encountered by developers when using Web APIs. As difference with our work, they put all posts (from both Web APIs and \sose) in a same dataset, which is later used for extracting topics using LDA.
Lee et al. \cite{Lee2017} studied the similarity in developer interests within and across GitHub and \sose. They found that the 39\% of the GitHub repositories and \so questions that a developer had  participated fall in the common interests.
Zagalsky et al. \cite{Zagalsky2016:RCurates} focused on R language by analyzing questions and answers  from two channels: the R-tag in \so and the R-users mailing list. They found  that knowledge is constructed in each channel in a different manner:
on \so participants contribute knowledge independently of each other, whereas R-user mailing list are more likely to build on other answers.
In this work, we create two datasets of \qa related to Xamarin technology, that could be used for replicating some of these works (e.g., \cite{Zagalsky2016:RCurates}).

\begin{comment}
\paragraph{Apps Stores Analysis}

Several works have focused on studying mobile application stores such as Google Play as presented by the survey of  \cite{surveyappstores}.
In our best knowledge, some works analyzed the reviews and rankings of cross-platforms mobile applications \cite{Ali2017SAD,hu2016crossconsistency,Malavolta2015Hybrid}.
For example, 
Ali et al. \cite{Ali2017SAD} detected 80,000 cross-platforms apps  from a corpus of 2.4 million apps collected from the Apple and Google Play app stores.  
They found that 80\% of the top-rated apps are cross-platforms and that the Android version of app-pairs receives higher user-perceived ratings compared to the iOS version.
Malavolta et al. \cite{Malavolta2015Hybrid} have focused on analyzing hybrid mobile apps (i.e., those developed using frameworks such as Apache Cordova/PhoneGap) available on the app store Google Play. They mined \numprint{11917} free apps and their related metadata from that store, finding is that the average of end users ratings for both hybrid and native apps are similar (3.75 and 3.35 out of 5, respectively). 

\end{comment}

Other works focus on classifying, comparing and evaluating  cross-platform mobile application development tools  to build  hybrid mobile  and native apps (\cite{heitkotter2012evaluating,francese2013supporting,dalmasso2013survey,palmieri2012comparison,desruelle2012challenges}). 
For example, Ciman et al. \cite{Ciman2016} analyzed the energy consumption of mobile development:  
their results showed the adoption of cross-platform frameworks as development tools always implies an increase in energy consumption, even if the final application is a real native application. In that work, Xamarin framework was not evaluated.
To the best of our knowledge, no work has studied cross-platforms mobile applications created with Xamarin framework.
In this work we study questions from \qa to understand the main problematic faced by developers when they develop cross-platforms mobile apps using Xamarin framework.

\section{
Creation of two datasets with  Q\&A about Xamarin
}
\label{sec:datasets}

In this section, we present two datasets with questions and answers (Q\&A) related to Xamarin technology.
In  section \ref{sec:sodataset} we present the first dataset  that is composed of posts extracted from the site Stack Overflow.\footnote{\url{https://stackoverflow.com/}}
In section \ref{sec:xamforum} we present the second dataset, named \xfdbse, is composed of posts extracted from the official forum of Xamarin.\footnote{\url{https://forums.xamarin.com/}}
Finally, in section \ref{sec:analyzingqa} we briefly describe the data from both datasets.
In section \ref{sec:results} we use both datasets for extracting the main discussion topics related with Xamarin technology.

\subsection{Dataset of \qa from Stack Overflow}
\label{sec:sodataset}

\subsubsection{Data Description}
We used the data dump of Stack Overflow provided by Stack Exchange, Inc.\footnote{\url{https://archive.org/details/stackexchange}}
The data is organized on 8 XML files, each one represents one data entity: 
 {Posts}, {Users}, {Comments}, {Tags}, {Votes}, {PostLinks}, {PostHistory}, and  {Badges}.\footnote{\url{https://ia800500.us.archive.org/22/items/stackexchange/readme.txt}}
For facilitating the manipulation of the data, we migrated it to a relational dataset.\footnote{\url{https://gist.github.com/gousiosg/7600626}}

A post from the \so data dump corresponds to a question or an answer done by an user.
Question can have zero or more tags (up to 5).
A post can also have a vote, which represents a mark as `favorite', `spam', `inform moderator', `offensive', etc.
The \so dataset contains \numprint{14458875} questions and 
\numprint{22668556} answers.\footnote{Data dump released the August 31, 2017}

\subsubsection{Methodology to filter Xamarin posts from Stack Overflow}
\label{sec:detectingxampostinso}

The first  challenge is to filter posts related to Xamarin technology among all posts present in the \so data dump.
For that, we apply two techniques.
The first one filters posts according to the tags associated to each post. 
The second technique matches keywords from the posts' titles.
Let us detail each strategy.

\paragraph{Filtering posts by tags}
\label{sec:tags}

We used the technique proposed by \rosen \cite{Rosen2016MDA}, who filtered posts related to mobile technologies.
Their technique consists of three steps.
The first step filters posts using a initial set of tags (in that work the author used mobile related tags such as `Android', `iOS'). 
The second step analyzes the most representative and significant tags from the  posts retrieved after applying the first step.
Finally, the last step filters posts from \so using the most representative tags for the mobile technology (those tags discovered in the previous step).

\begin{table}[t]
\caption{Tags used for detecting questions related to Xamarin. The technique for filtering tags is described in  \cite{Rosen2016MDA}. The table shows the total occurrence of the tag in all posts, in posts tagged with 1+ tag from the final set of tags, and two measures TRT and TST used by the technique for filtering tags. The table shows the ten tags with highest number of posts.}
\begin{center}
\scriptsize
\begin{tabular}{|l|r|r|r|r|} 
\hline
\multirow{2}{*}{Tag Name} & \multicolumn{2}{|c|}{ Occurrences }  & \multirow{2}{*}{TRT} & \multirow{2}{*}{TST}\\ 
\cline{2-3}
 &All posts& Xamarin posts  & &\\ \hline
\hline
xamarin & 25029 & 25029 & 100\% & 100\% \\ \hline
xamarin.ios & 10640 & 10640 & 100\% & 42,51\% \\ \hline
xamarin.android & 9590 & 9590 & 100\% & 38,32\% \\ \hline
xamarin.forms & 9308 & 9308 & 100\% & 37,18\% \\ \hline
mvvmcross & 2881 & 2003 & 69,52\% & 8,00\% \\ \hline
xamarin-studio & 1355 & 1355 & 100,00\% & 5,41\% \\ \hline
monodevelop & 2430 & 696 & 28,64\% & 2,78\% \\ \hline
portable-class-library & 1355 & 502 & 37,05\% & 2,01\% \\ \hline
monotouch.dialog & 461 & 428 & 92,84\% & 1,71\% \\ \hline
xamarin.mac & 386 & 386 & 100,00\% & 1,54\% \\ \hline
...& ... & ...  & ...  & ...  \\ \hline
\end{tabular}
\end{center}
\label{tab:tags-metrics}
\end{table}

We applied the technique as follows.
In the first step we defined the initial set of tags:  
as we aim at filtering Xamarin-related posts, our initial set was ``\%xamarin\%'', where \% corresponds to zero or more characters. 
In total, 28 tags include the word `xamarin' such as `xamarin.iso', and formed the initial set of tags.
We then retrieved \numprint{39855} questions tagged with at least one tag from the initial set. 
Those \numprint{39855} questions have, in total, \numprint{4143} tags.
Note that, those retrieved posts could also be tagged with tags that are not included in the initial set, but they are related to Xamarin technology such as `monodevelop'. 
To obtain tags related to the Xamarin technology not included in the initial set, the technique by \rosen  presents two measures, TRT (tag relevance threshold) and TST (tag significance threshold) \cite{Rosen2016MDA}, that help us to:
\begin{inparaenum}[\it a)]
\item obtain those tags representative to Xamarin technology, and 
\item discard those tags that are related to Xamarin technology but, at the same time, they are too general.
For instance, a significant portion of  Xamarin posts are tagged with "Android". However,  most of the questions tagged with Android are not related to Xamarin technology.
\end{inparaenum}

TRT($tag_i$) is the ratio between the
number of posts that contain at least one tag from the initial set and the  total number of posts related to $tag_i$.
TST($tag_i$) is the ratio between the number of mobile posts for $tag_i$ and the number of mobile posts for the most popular mobile tag.

In the second step, we filtered those tags with at least a 25\% for representative (TRT) and 0.1\% for significance (TST).
In total,  \numprint{38}  tags conforms the final set of tags related to Xamarin technology.  
Table \ref{tab:tags-metrics} shows  the 10 tags with higher number of posts, and for each tag 
the number of total occurrences on a) all posts and b) on Xamarin posts, and the measures TRT and TST. 
For example, 
the final tag set includes tag "monotouch.dialog", which has a TRT of 92.84\%:
almost all posts tagged with that tag  are also tagged with one tag from the initial tag set such as `Xamarin'.
Moreover,  those posts represent the 2.34\% (i.e., the TST value) out of all posts tagged with at least one initial tag.
On the contrary,  the tag `Android' is not included in the final tag set: its TRT metric is lower that the threshold we set:  
only the \numprint{1,07}\% of post tagged with Android also are tagged with Xamarin (i.e., TRT metric).

Finally, for the third step of the technique, we retrieved \numprint{43988} posts related to Xamarin technology:
\numprint{39855} of them tagged with the initial tags (i.e., include the keyword "Xamarin") and  \numprint{4133} posts  tagged with at least  one tag from remaining final set of tags. 

\paragraph{Filtering posts by keywords}
\label{sec:keywords}

The second technique for retrieving Xamarin-related posts filters posts by matching \emph{keywords} on the post's title.
This strategy aims at retrieving those posts that are not tagged with any tag from the final set of tags.
We define a keyword for each tag included in final set of tags presented before.

Using this keyword-based strategy, we retrieved \numprint{446} new posts related to Xamarin technology.
For instance, one of them is the post  "Xamarin Android Save sms" (post id 29405420)\footnote{\url{https://stackoverflow.com/questions/29405420/xamarin-android-save-sms}}
 which was tagged with 3 tags: "C\#, android, datetime", none of them included in our final set of tags.

\paragraph{Final set of Xamarin-related posts from \sose}

Using the two described techniques, we retrieved \numprint{44434} questions from the \so dump:
\numprint{43988} were found using the tag-based strategy, whereas
\numprint{446} were found using the keywords-based strategy.

\subsection{\xfdbse: a dataset with questions from \xf}
\label{sec:xamforum}

\subsubsection{Describing Forum Xamarin}
\label{sec:descxamforum}

The \xf is an online web platform where mobile developers post questions or start a discussion about the Xamarin framework and its ecosystem. 
Moreover, it is a communication channel between the developers of the framework (a.k.a. Xamarin Team) and users (i.e., the developers of mobile apps that use Xamarin).
For instance, new versions of the framework or of particular components are announced in the forum.\footnote{\url{https://forums.xamarin.com/discussion/85747/xamarin-forms-feature-roadmap\#latest}}
All posts are publicly available. However, for creating a new post, users must be registered into the site, which is free.

\paragraph{Forums}
The \xf site is composed of seven general forums:
Community, General, Pre-release \& Betas, Tools and Libraries, Graphics \& Games,
Xamarin Platform and Xamarin Products.
For instance, the forum Xamarin Platform contains questions related to the development of an application for different targeted platforms, while the forum Xamarin Product focuses on discussions about products related to the Xamarin technology such as TestCloud, a cloud-based platform for testing mobile apps built using Xamarin.
Each of those forums has one or more `sub-forums'.
For instance, the mentioned Platform forum has 5 categories: Android, iOS, Cross-Platform, Mac, and Xamarin.Forms.
The forums are created by the forum administrators, which means that users are not able to define new ones.

We identify two types of forums: 
\begin{inparaenum}[\it 1)]
 \item those related to technological topics (platforms, libraries, tools, IDEs, etc.)
and 
\item those related to non-technological topics related to Xamarin (events, conferences, jobs, etc.).
\end{inparaenum}

The main page of each forum shows a paged-list of posts and two buttons, one for creating a question and the other for creating a discussion. 
Each post from the list shows the post's title, author, number of views, number of answers, and zero or more labels.
Those labels are colored boxes located near the title and indicate, for instance, if a question was answered or if the answered was accepted by the user who wrote the question. 

\paragraph{Posts}
A user can create a post on a given category.  
Even there are two types of posts, i.e., questions and discussions, the creation forms of them are similar, i.e., both have the same fields.
The user can also select a list of tags associated the new post. 
However, once a post is created, the forum does not show the list of tags associated to a post.\footnote{Last visit: {June, 2018}}

\paragraph{Answers}
A registered user can answer an existing question or to put a `like' on existing answers.
Moreover, as in others Q\&A such as Stack Overflow, the post's author can \emph{accept} one or more answers. The accepted answers are labeled with a green-colored box tag ``Accepted answer".

\paragraph{Users}
The \xf provides the forms  for registering new users, which have the right to create new posts and to write comments.
In the forum also participate users that belong to the organization that develops Xamarin framework.

\subsubsection{Dataset construction}

The Xamarin Forum site does not have an API to programmatically access and retrieve the data, such as those provided by Stackoverflow\footnote{\url{https://api.stackexchange.com/}} and GitHub\footnote{\url{https://developer.github.com/v3/}}.
For this reason, we developed a web crawler for retrieving and storing all pages from \xfse. 
Our engine has main two phases: 
\begin{inparaenum}[\it 1)]
\item Page fetching and 
\item Page parsing.
\end{inparaenum}

\paragraph{Fetching pages}
\label{sec:fetching}
The first phase fetches (i.e., downloads)  all web pages written in HTML from the forum web site \url{https://forums.xamarin.com/}, by  accessing  via HTTP protocol.
The \xf has two kinds of web pages. 
A page from the first kind corresponds to a single post.
It contains the post's title, question (or discussion topic), author data (names, location, roles), posting date and a list of comments. Each comment has the name of the user that wrote it, the date, and one or more labels such as ``Answered question".
The second kind of pages corresponds to the main pages of each forum. 
A ``main" page shows in a paged list all posts done in that forum\footnote{In those main pages, those posts are called `Threads'}, ordered by decreasing creation date.
The list shows for each post its title, the numbers of views (i.e., number of visits the post has received), numbers of comments done, and zero or more labels that indicates if the post is a question or announcement, if it has an accepted answer, etc.

\paragraph{Extracting data from pages}

In the second phase, our engine extracts and format data for each fetched page and store it in a  relational database.
Our engine is implemented in python language, and uses the library BeautifulSoup\footnote{\url{https://www.crummy.com/software/BeautifulSoup/bs4/doc/}} for parsing the Xamarin web pages in HTML.  The structured data is then stored in a MySQL database.

\subsubsection{Data schema}{\label{sec:dataschema}}

The store all  data extracted from the Xamarin forum in a relational database.
We now describe briefly the main entities from the database schema. 
The first entity is \ent{Post}, which stores all posts written in the Xamarin forums. Each post belongs to one \ent{Forum} and has zero or more \ent{Comments}. 
Note that the post entity stores both questions and discussions.
We decide to model both questions and discussion in the same entity due they share most of their properties.
Both \ent{Comment} and \ent{Post} entities have a relation {\it many-to-one} with the entity \ent{User}, which stores the information of the users that make or answer a post.
A \ent{User} has different \ent{Roles}. 
A registered user has the role `Member' by default, but there are other roles exclusively assigned to used belonging the Xamarin organization such as `Forum Administrator', or role `XamUProfessor" which are professionals involved on the Xamarin training program called Xamarin University.\footnote{\url{https://www.xamarin.com/university}}

\subsection{Perspectives of \xf and Xamarin-related posts from \sose} 
\label{sec:perspectivexamdb}

In our opinion, both datasets can be used by the research community for  understanding the main concerns about developing cross-platform mobile applications using Xamarin framework.
In this paper  we use them to discover the main discussed topics from their questions.

\section{Analyzing  datasets  with \qa related to Xamarin technology}

\label{sec:analyzingqa}

In this section we describe and compare the two datasets with questions and answers related to Xamarin technology presented in section \ref{sec:datasets} with the goal of answering the first research question: {\it \questionamount}.

\subsection{Number of questions}

The   \xfdb has \numprint{91838} questions.\footnote{In this paper, we use the terms `question', `post', and `thread' interchangeably.}
The first dates from January, 29 2013, and the last one from September, 6 2017.
There are \numprint{85908} (93.5\%) questions related to technological forums,  and \numprint{5930} (6.5\%) related  non-technological forums such as Events has \numprint{1079} questions (1.17\%), Job Listings 465 (0.51\%).
In this work, we are interested on those technological questions, discarding the non-technological for further analyses. 

In \so we selected \numprint{44434} questions discussing about Xamarin technology. 
They represent around the 0.12\% of all questions from \sose, which has \numprint{37215528} questions.
Compared to other \xpmf, the number of questions has the same order of magnitude.
For instance, \numprint{21515} questions were tagged with "\%react\%native\%" (framework for generating native mobile apps using JavaScript language),
\numprint{60790} questions were tagged with "\%cordova\%" and  \numprint{8194} "\%phonegap\%" (Apache Cordova/Phonegap, is a framework for creating hybrid mobile apps using HTML and JavaScript).\footnote{Data corresponding to dump from August 31, 2017.}

\subsection{Number of answers and acceptance}
\label{sec:nranwsers}

On \xfse, the 72.3\% (\numprint{62114} out of \numprint{85908}) of the technological questions have at least one answer. In average, each question has 2.99 answers. The remaining 29.7\% (\numprint{23794}) has not been answered.
Moreover, the 15.2\% (\numprint{13025} out of \numprint{85908}) of the questions have, at least, accepted answer.
The proportion of answered questions is similar in \so and \xfse: 79\% vs 72.3\%.
However, the proportion of accepted answers on \so is 3 times as larger as that one from Xamarin forum: 46.1\% vs 15.2\%.

On \sose, the 79\% (\numprint{35100} out of \numprint{44434}) has at least one answer. 
In average, each question has 1.13 answers.
The  46.1\% (\numprint{20495} out of \numprint{44434}) of Xamarin questions has one accepted answer,
whereas the 33.9\% (\numprint{14605}) has at least one answer but any on them was accepted.
The remaining 20\% (\numprint{9334}) has not been answered.

\subsection{Number of views per question}
\label{sec:nrviews}

Xamarin-related questions from Stack Overflow were viewed \numprint{36523008} times, 821.9 views per question as average, whereas 
those from \xf were viewed \numprint{35032568}, that is, 407.8 views per question as average.
In conclusion, questions from \xf have received almost the same number of views than Xamarin-related questions from \so (>35 millions).
However, \so questions are viewed, in avg., two times more that those from Xamarin forum. 

\begin{myframe}
{\bf Response to {{\rqamount}}}
\begin{inparaenum}[\it a)]
\item Xamarin forums has a 85,908 questions related to Xamarin and  Stack Overflow has 44,434.
\item  The proportion of answered questions is similar in Stack Overflow and Xamarin Forum (79\% vs 72.3\%), however, the proportion of accepted answers on Stack Overflow is $\sim$3 times as larger (46.1\% vs 15.2\%);
 \item Questions of both sites  have received almost the same number of views (>35 millions), however, those from Stack Overflow are viewed, in avg., two times more.
\end{inparaenum}
\end{myframe}

\section{Replication study over the two datasets: topic modeling}
\label{sec:replication}
In this section, we present an study that replicates the experiment replicates the study of \rosen  \cite{Rosen2016MDA}. That work  focuses at discovering topics from mobile development (i.e., not related to neither a particular platform nor framework) from  \sose.
On the contrary, our study aims to discover topics from Xamarin-related questions from  \so and \xfse.
In addition, our replication study compares those Xamarin-related topics with those found by \rosense.
This section is organized as follows:
In Section \ref{sec:methodology} we first introduce the methodology to discover topics (based on that one from \rosense), and in Section \ref{sec:results} we then present the results.
In the rest of this paper, when the mention \so questions we refer to the dataset of questions from \so related to Xamarin technology that we built in section \ref{sec:sodataset}, whereas \xf questions are those from the \xfdb built in section \ref{sec:xamforum}.

\subsection{Methodology for replicating an experiment about topic modeling}
\label{sec:methodology}

In this section, we present three methodologies that we execute to respond the research questions.
First, in section \ref{sec:protocolMainTopics} we present a methodology  for discovering the discussion topics from questions from Stack Overflow and  Xamarin Forum (i.e., from the datasets we presented in sections \ref{sec:sodataset} and \ref{sec:xamforum}).
The results will us allow to respond the second research question. 
Then, in section \ref{sec:methodmatching} we present a methodology for relating topics discovered from different sources. We use it to respond the third  research question.
Finally, in section  \ref{sec:metodol_relevant}  we introduce a methodology for retrieving relevant questions from a topic and we use it to respond the forth research question.

\subsubsection{Methodology for Discovering Topics}
\label{sec:protocolMainTopics}

This section presents a methodology for discovering the main topics from a dataset of questions. We applied this procedure for obtaining two sets of main topics:
one from \so questions, another from \xf questions.

The methodology consists of the following steps.

\paragraph{Retrieve posts titles} 
For each question from a dataset (Xamarin Forum or \sose), we created a \emph{document} that includes all words contained in the question's title as done by \rosen  \cite{Rosen2016MDA}.
In case of \xfse, we only considered the technological posts (i.e., the 93.5\% of all posts), as discussed in Section \ref{sec:detectingxampostinso}.  

\paragraph{Document processing} 
We pre-processed all documents by first removing stop words (e.g., "at", "the") and then applying a customized word stemming processing.
During the setup of our experiment, 
we noted that traditional stemming (as applied by, for example, \cite{Linares-Vasquez2013EAM})  alternated domain-specific  words  such as ``iOS" or ``Mono", and, by consequence, the step produces a loss of information.
With the goal of minimizing the impact of traditional stemming,  
we defined a   \emph{customized} word stemming process composed by the following steps.
We first created a list of words related to the Xamarin technology. Those words are the tags related to Xamarin posts presented in section \ref{sec:tags}.
The list, called \emph{technology-specific words list},  is available in our appendix.
Words from a post title that belong to that list are not stemmed and they are included in a document without any transformation applied.
For the words that are not included in the list, 
we applied the first step of the stem process described by \cite{Porter1997ASS} for transforming all words in singular.

\paragraph{Discovering topics}
We applied Latent Dirichlet allocation (LDA) algorithm as done by Blei et al. \cite{Blei2003LDA} for discovering topics from a set of documents, where each document corresponds to a pre-processed question's title.
Several works have used LDA for discovering main topics in the domain of, for example,  web development \cite{Bajaj:2014}, software maintenance \cite{Sun2015MSR}, and mobile development (\cite{Rosen2016MDA,Linares-Vasquez2013EAM}).
As output, LDA produces a set of topics, each composed of a list of pairs word-probability. Typically, previous works represent a topic with the  15 or 20 words with highest probability (\cite{Linares-Vasquez2013EAM}  and \cite{Rosen2016MDA}, respectively).
We used an implementation of LDA called Mallet  \cite{mccallum2002mallet}.

One of the challenge of LDA to chose a configuration (i.e., values for the input parameters) that produce meaningful topics.
There are four main parameters to configure: 
\begin{inparaenum}[\it a)]
\item alpha,  
\item beta,  
\item number of topics to generate, and 
\item number of iterations.
\end{inparaenum}
In this experiment, we used a similar configuration proposed by \rosen \cite{Rosen2016MDA}, one of the closest related work, which studied what mobile developers are asking about on \sose. 
The configuration they used is:
40 topics to generate (nrtopics),  alpha = 0.025 (that is: 1/nrtopics); beta = 0.1 and number iterations = 1000.
Having similar configuration to \rosen allows us to compared the topics we discovered from Xamarin-related questions, with those from native mobile apps discovered by them \cite{Rosen2016MDA} (section \ref{sec:topicsrosen}).

In the remaining of the paper, 
we refer as $t_{so}$ $N_i$ and $t_{xam}$ $N_i$, to topics with identifier $N_i$ from \so and Xamarin, respectively.

\paragraph{Labeling topics}
\label{sec:labelingtopics}

As each topic discovered from LDA is a list of words,
previous works labeled each topic with a human readable label.
In this work we decided to reuse, when is possible, labels related to mobile technology defined by \rosen  \cite{Rosen2016MDA}, which discovered and labeled topics related to mobile development.
Some of the labels from this work are ``Input'', ``Tools'',``Application Store'',  ``Parsing'', ``Map and Location''.
In case that  none of their  labels describe correctly a topic that we discovered, we defined a new label by taking into account the words from the topic and their probabilistic.
In some cases we used the same label for two related topics and included a sub-category in parenthesis for remarking the differences between them.

\subsubsection{Methodology for relating topics from different data sources}
\label{sec:methodmatching}
In our experiment, we need to find whether a topic from a dataset (e.g., \xfse) is also a topic from another dataset (e.g., \so or that one from \rosense).

We related two topics from different datasets if they have:
\begin{inparaenum}[\it a)]
\item similarity on their topic labels;  or
\item similarity on the most significant  works that conform the topic.  The significant of a word is the  probability given by the generated LDA model to that word.  
\end{inparaenum}
For example, we related topics $t_{so}$  28  with $t_{xam}$ 40 because they have the same topic label: ``User Interface (Table)''.
As the labels do not always match, we related other topics using the words from the topic. 
For example, topic $t_{so}$ 8 was label as ``Graphic and memory", and the most similar topic in Xamarin that we found is $t_{xam}$ 16, which was labeled as ``Video memory''. 
We related both topics because they shares significant words like ``memory'' and ``leak''.
The lists of topics (words and their probabilities) and the related topics of each are available in our appendix.

\subsubsection{Methodology for filtering most relevant questions from a topic}
\label{sec:metodol_relevant}

In this paper, we define  \emph{relevant questions} of a topic $t$  as those questions $qs$ from  $t$ with:
\begin{inparaenum} [\it a)]  \item most number of views, or 
\item highest score, i.e., number of up votes (only for \sose).
\end{inparaenum}
Note that all questions we consider for analyzing topic $t$ have $t$ as dominant topic. 
A \emph{dominant topic} $t$ of document $d$ is the most related topic (i.e., with highest probabilistic) according to the LDA topic  model generated.

The intuition is that if a question asks about a relevant subject or a recurrent problem about Xamarin technology, it is frequently visited and appreciated with up votes done by other developers.
For each topic, we selected the questions with at least \numprint{10000} views and 10 up votes as score.

\subsection{Results:}
\label{sec:results}

\subsubsection{\rqtopics
} 
\label{sec:maintopicresults}

\begin{table}[t]
\caption{Main Topics discovered from \so (Left) and Xamarin (right).
}
\centering
\scriptsize
\begin{tabular}{|r|r|r|r|}

\hline
\multicolumn{2}{|c||}{\so}&\multicolumn{2}{|c|}{Xamarin}\\
\hline
\hline
Id & Label & Id& Label \\
\hline
\hline
11,32	&	List (Forms)	&	7,10	&	View Controllers/Navigations	\\
28	&	User Interface (Table) 	&	4,8	&	Lists (Forms)	\\
23	&	User Interface (Layout)	&	3	&	Forms (WebView) 	\\
22	&	Emulator/Device/Simulation	&	1	&	Resources/File	\\
3	&	Web Services  	&	5	&	Android (Activity)	\\
5	&	Forms (XAML)	&	17	&	User Interface (Style)	\\
1	&	Monodevelop/.net framework	&	11	&	Cross-Platform  	\\
30	&	View Controllers/Navigations	&	2	&	IDE/Licence	\\
19	&	MVVM	&	19	&	Forms (XAML)	\\
4	&	Library/Native 	&	6	&	Map \& Location	\\
10	&	HTTP Request	&	28	&	IDE (Visual Studio)	\\
7	&	Packages/Nuget/PCL	&	21	&	Social/APIs	\\
21	&	Inputs (Event)	&	32	&	IDE (Simulator- for-Mac)	\\
20	&	Android (Activity)	&	15	&	Application Store	\\
39	&	User Interface (Style)	&	35	&	User Interface (Layout)	\\
13	&	Library/Portability	&	18	&	Notifications	\\
24	&	Databases	&	34	&	Emulator/Device/Simulation 	\\
8	&	Graphics/Memory	&	20	&	Library/Native	\\
17	&	IDE	&	25	&	Inputs (Text)	\\
9	&	Language Questions 	&	26	&	Media/Images 	\\
18	&	Android (Debug/Device)	&	9	&	Error (Unified/Insight) 	\\
2	&	Exception/Error	&	12	&	Android (Version)	\\
12	&	Phone Orient./Media-images	&	30	&	Databases	\\
25	&	Error (Code/Compilation)	&	31	&	PCL/Library	\\
6	&	Resources/File	&	14	&	Nuget/Package	\\
29	&	Monodevelop	&	37	&	Web Services 	\\
35	&	Social/APIs	&	24	&	Inputs (Events)	\\
27	&	File Operations	&	16	&	Video/Memory	\\
26	&	Media/Images	&	13	&	Releases	\\
16	&	Notifications	&	40	&	User Interface (Table)	\\
15	&	User Interf.(Forms/WebView)	&	38	&	Connectivity	\\
14	&	Media/Streaming /Video	&	22	&	Testing	\\
33	&	Location	&	33	&	Language Questions (OOP)	\\
31	&	Input (Text)	&	36	&	Phone Orientation/Media-images	\\
34	&	Threading 	&	27	&	Questions (Forms, Samples)	\\
36	&	Error Assembly	&	23	&	Exception	\\
40	&	Binding/library	&	39	&	ErrorAssembly	\\
38	&	Windows Platform	&	29	&	Error (Code/Compilation)	\\
37	&	Testing	&		&		\\

\hline
\end{tabular}
\label{tab:mainTopicsXamSO}
\end{table}
 
Table \ref{tab:mainTopicsXamSO} shows the topics from \so and \xf questions discovered using the methodology described in section \ref{sec:protocolMainTopics}.
The left-size displays the \so topics and the right-side displays the Xamarin topics.
The column \emph{Id} corresponds to
the topic identifier assigned by Mallet; the column \emph{Label} corresponds to  the  label manually assigned by us using the method presented in section \ref{sec:labelingtopics}.
The topics are sorted by decreasing NDDT. As explained by Linares-Vasquez et al. \cite{Linares-Vasquez2013EAM} NDDT counts for each topic $t$ the number of documents that has as $t$ as \emph{dominant} topic. 
In our appendix we presented,  for each discovered topic,  the words that compose the topic, their probabilities, and NDDT metric.
Let us first present some of the discovery topics, and then to compare the topics discovered from both datasets. 

\paragraph{Main Discussion Topics Discovered}

Let us first focus on \sose:
the top-4 topics (according with the number of dominant documents NDDT) are related related to the view: 
$t_{so}$ 11 and 32 both labelled with "List (Forms)";
$t_{so}$ 28 labelled with "User Interface (Tables)";  and 
$t_{so}$ 23 "User Interface (Layouts)".
Then, the next two topics are not directly related to the user interface, for example: $t_{so}$ 22 "Emulator/Device/Simulator"
and $t_{so}$  3 "Web Service".

For \xfse, the top-5 topics from \xf are related to View or Controllers development:
$t_{xam}$ 7 and $t_{xam}$ 10 both labelled with  "View Controllers/Navigation",
$t_{xam}$ 4 and $t_{xam}$ 8 both labelled with "Lists (Forms)", 
and 
$t_{xam}$ 3 labelled with "Forms (WebView)".
Then, after them there are two topics not related to the user interface:  $t_{xam}$ 1 "Resources/Files" and $t_{xam}$  5 "Android" (Activity).

When we analyzed the remaining topics, we observed that the topics discussed in the \qa are diverse.
For instance, those cover topics such as:
resources and files  ($t_{so}$ 6   and  $t_{so}$ 1),  web services ($t_{so}$ 3 and $t_{xam}$ 37), language questions ($t_{so}$ 9  and $t_{xam}$ 33),  databases ($t_{so}$ 24 and topic$_{xam}$ 30), notifications ($t_{so}$ 16 and $t_{xam}$ 18), packages ($t_{so}$ 7 and $t_{xam}$ 31), 
architectural  ($t_{so}$ 19,  $t_{xam}$ 19), social ($t_{so}$ 35 and $t_{xam}$ 21), maps and locations ($t_{so}$ 33 and $t_{xam}$ 6), IDEs ($t_{so}$ 17 and $t_{xam}$  28).

\paragraph{Common Main Topics \so and \xf \qa sites}
\label{sec:commonXamTopics}
 
We observe that \xf and \so  share 33 out of 40 (82.5\%) of the discovered topics.  
For instance, there are questions that discuss about ``Notifications'': in \so those questions have as dominant topic $t_{so}$ 16 and those from \xf has as dominant topic $t_{xam}$ 18. 
Seven topics from Xamarin and other 7 from \so were not related to any topic.

\paragraph{Main topics found from only one \qa site}

In both \so and \xfse, there are 7 discovered main topics (12.5\%) that we could not manually related to any topic of the other \qa site.
For instance, we discovered that one of the main topic from \sose, $t_{so}$ 34, is about  ``Threading''.  However, we could not find  any  topic between those from Xamarin that discusses about threading. 
On the contrary, we discovered  from the \xf site a topic ($t_{xam}$ 15)  which discusses about Application Store, but none of the main topics from \so is related to it. The lists of unrelated topics are available in our appendix.

\begin{myframe}
{\bf Response to \rqtopics}
The top main topics discovered from \so and Xamarin forum discuss about User interfaces: Tables, Layouts, Controllers, Forms.
The 82.5\% of the discovered main topics are present in both \so and \xf sites.
\end{myframe}

\subsubsection{\rqnativet}
\label{sec:topicsrosen}

As in this paper we have replicated the study of \rosen (which focuses on general mobile development) over a Xamarin-related \qa datasets, we now proceed to compare the results  (i.e., discovered topics) of both studies.
In this way, we are able to detect  topics discovered  from \xf and \so 
that are:
\begin{inparaenum}[\it a)]
\item \emph{general}, i.e.,  related to mobile but not particularly to Xamarin technology,  and
\item  closely related to the development of cross-platform mobile applications using Xamarin.
\end{inparaenum}

\paragraph{Comparing topics from different \qa datasets}

We compared the topics previously discovered in section \ref{sec:maintopicresults} with those  related to  mobile application development presented by \rosen \cite{Rosen2016MDA}. 
As the authors analyzed questions about three mobile platforms (Android, iOS, and Windows Phone), we denominated their topics as `general', and we reference each as  $t_{genmob}$ N, where N is the id of the topic.\footnote{As the topics from \cite{Rosen2016MDA} do not include any `id', we consider that the ids of them correspond to the row numbers of the table that presents the topics.}
For comparing topics, we used the same manual methodology for matching topics presented in section \ref{sec:methodmatching}.
Note that, with the goal of carrying out fair comparison,  for discovering  topics from a corpus of questions, we used the same technique (LDA) and configuration (values for alpha, beta, \# iterations and \# topics) that  \rosen used in their work.

We found that 27 topics from \so  and 30 topics from \xf are related to, at least, one \emph{general} topic from \rosense.
For instance, 
topic $t_{so}$ 35 is related to  $t_{genmob}$ 8 due to both share the same label: ``Social/APIs''.
Other topics were also related by using the topics' words instead of their labels.
For instance, topic $t_{so}$  21 labeled as ``Inputs (Event)'' was related to  $t_{genmob}$ 1 labeled as  ``Input'' (label more general than the previous one), due to both topics  have almost the significant words (i.e., those with higher probability): `button', `event',  `click', `android' and  `keyboard'.

\medskip
 
\begin{myframe}
{\bf Response to \rqnativet}
There are 27 and 30  topics  discovered from \so and \xf (67.5\% and 75\%, resp.)  that are also main topics in question about mobile application.
The remaining topics, i.e., 13 (32.5\%) and 10  (25\%) topics from  \so  and \xfse, resp., are particularly related to cross-platform app development using Xamarin.
\end{myframe}

\bigskip

This overlap between Xamarin and general topics makes sense since:
\begin{inparaenum}[\it a)]
\item Xamarin framework produces general code for different platforms, which it could be  maintained during the app life-cycle in the same way as any general app developed using traditional tools.
\item  a portion of a Xamarin application is usually written in general language (Java for Android, Objective-C or Swift for iOS) rather than in the common language (Xamarin uses C\# as common language).
for example, the Evolve application for Android platform has the 90\% of code written in the C\# whereas the remaining 10\% in general code.\footnote{\url{https://github.com/xamarinhq/app-evolve}}.
Thus, in both cases, when writing or maintaining the portion of general code, a Xamarin developers cold have the same kinds of questions than developers of general apps.
\item the general development of mobile apps includes the development of cross-platforms mobile apps. 
\end{inparaenum}

Now, let us to analyze those main topics that are not include between the general topics, and those general main topics that are not present between the main topics discovered from \so and \xfse.

\paragraph{Main topics from \so and \xf sites about Xamarin not previously reported}

Let us start discussing the discovered topics from \so and \xf not reported by \rosense.
There are 13 and 10 discovered topics from \so and \xfse, respectively, that could not be related to any general  topic. 
For instance, the topic $t_{so}$ 19	labeled as ``MVVM'' is one of them.
As its label indicates,  $t_{so}$ 19 represents documents that discuss about the design pattern Model–view–viewmodel (MVVM), which was introduced by Microsoft for facilitating the design of multi-tiers application under the  Microsoft's platforms .NET.
This pattern is recommend by Xamarin documentation for implementing large and complex applications on Xamarin platform.\footnote{\url{https://developer.xamarin.com/guides/xamarin-forms/enterprise-application-patterns/mvvm/}}
Another topic is  $t_{so}$ 7 ``Packages/Nuget/PCL'' which refers to concepts from the Microsoft technology.
The first one is "Portable Class Library (PCL)", which is a type of project in the .NET framework to write and build portable .NET assemblies that are then referenced from, for example, cross-platform apps.
\footnote{\url{https://docs.microsoft.com/en-us/dotnet/standard/cross-platform/cross-platform-development-with-the-portable-class-library}}
The second concept is ``Nuget'', a package manager for .NET.\footnote{\url{https://www.nuget.org/}}
In summary, it makes sense to find this topic only Xamarin-related questions from \so and \xfse: it covers questions about reuse of functionality under the .NET platform during the development of cross-platform mobile application using Xamarin framework.
Similarly, in \xf we discovered two topics, each related to one mentioned concept:	$t_{xam}$ 31 ``PCL/Library''  and  $t_{xam}$ ``14 Nuget/Package''.

Furthermore, we also discovered topics from  \so and \xf that are not neither related to any general topics nor directly related to Xamarin technology.
Two of them are topics  $t_{so}$ 37 and $t_{xam}$ 22 both labeled as ``Testing'', which include words such as  `uitest', `testing', 'unit', `test'. 
Note that no topic from \rosen discuses about testing: the mentioned words are not presented in any topic.

\paragraph{General topics not found in the Xamarin-related questions from \so and \xfse}

Between the general mobile topics discovered by \rosen about general mobile development,
there are 9 out of 40 that are not present in the main topics from \sose, whereas 10 are not related to any from \xfse.
7 of them could not be related to any topic from nether \so nor \xfse.
They are: 
 $t_{genmob}$  6 labelled with	``Phone/Sensors''; 
 $t_{genmob}$  12 ``HTML5/Browser''; 
 $t_{genmob}$  16 ``App Distribution''; 
 $t_{genmob}$ 	19	``Processes/Activities''; 
 $t_{genmob}$ 	20	``Data Structures''; 
 $t_{genmob}$ 	24	``Data Formatting'';  and
 $t_{genmob}$ 	30	``Contacts''.
 For instance, topic $t_{genmob}$  6 includes words that no discovered topic from \so and \xf has: as
`time', `alarm', `voice', `speech', `incoming', `number'.

\subsubsection{\rqquestions}
\label{sec:relevantquestions}

In section \ref{sec:topicsrosen} we discovered topics that discuss about Xamarin that were not previously reported by \rosen as main topics of general mobile development.
Now, we focus on three of  those topics to know the main concerts and problematic asked by Xamarin developers.
For each of them, we analyze the most relevant comments, according to the methodology presented in Section \ref{sec:metodol_relevant}.

\paragraph{MVVM ($t_{so}$ 19)}

The MVVM (Model-View-ViewModel) is a pattern that helps to separate the business and presentation logic of an application from its user interface (UI). 
The pattern was introduced by Microsoft for designing apps for its different platforms, including Xamarin, Windows Forms, WPF, Silverlight, and Windows Phone.\footnote{\url{https://msdn.microsoft.com/en-us/library/hh848246.aspx}}.

The most relevant questions from   topic MVVM are related to  Mvvmcross. 
Mvvmcross is a framework built for easing the development of Xamarin frameworks that proposes, for instance, an easier way to  implement the MVVM pattern.\footnote{\url{https://www.mvvmcross.com/}}
The question from \so  with highest score from topic  MVVM  (25 upvotes)  is  about asynchronous programming: 
\qt{How can I use async in an MVVMCross view model?} (id:17187113)\footnote{This number corresponds to the ID of a StackOverflow post. A post with id <ID> is publicly available at https://stackoverflow.com/questions/<ID> }. 
Between the answers, there are two that solve the problem: one (the accepted)  proposes to use the language keyword (\emph{async}), the other proposes a solution that relies on the Mvvmcross framework.

Other relevant questions about Mvvmcross focus on the  communication between the pattern's components, e.g.,
\qt{Passing on variables from ViewModel to another View [..]} (id:10192505),  
\qt{MvvMCross bind command with parameter [..]} (id:17492742). 
Moreover, another relevant question  is  about the differences between  Mvvmcross and ReactiveUI.\footnote{\url{https://reactiveui.net/}} (ReactiveUI is a model-view-viewmodel framework for all .NET platforms, including Xamarin).

\paragraph{User Interface (Forms/ViewList) ($t_{so}$ 11)}

Xamarin.Form is an API of Xamarin framework to build native apps for iOS, Android and Windows completely in C\# or in XML (using  XAML language).\footnote{\url{https://www.xamarin.com/forms}}
Xamarin.Forms pages represent single screens within an app, and  support  layouts, buttons, labels, lists, and other common controls.
Each page and its controls are mapped to platform-specific native user interface elements. 
Xamarin.Forms is best for developing apps that require:
\begin{inparaenum} [\it a)]
\item little platform-specific functionality, or 
\item code sharing is more important than custom UI.
\end{inparaenum}

The  relevant questions from this topic  $t_{so}$ 11 are about  Xamarin.Forms or its components.
For instance:
\qt{How to correctly use the Image Source property with Xamarin.Forms?} (id: 30850510) is the most viewed question.
Moreover, there are relevant questions about the component ListView,  e.g., 
\qt{[..] ListView inside StackLayout: How to set height?} (id: 24598261) and 
\qt{[..] untappable ListView (remove selection ripple effect)}	(id: 35586243).
The highest score questions of topic $t_{so}$ 11 are related to the IDE support  of XAML development, such as the code-completion (IntelliSense), e.g.:
\qt{Is it possible to use a Xaml designer or intellisense with Xamarin.Forms?} (id: 24158201) (36 up votes).
Other relevant questions are also about XAML, e.g.:
\qt{[..] ListView ItemTapped/ItemSelected Command Binding on XAML} (id: 24792991).

\paragraph{Library/Portability ($t_{so}$ 13)}

The topic Library/Portability  contains relevant questions related to the architecture of Xamarin-based cross-platform apps.
Xamarin provides three alternative  architectures that focus on  sharing code between cross-platform applications: 
\begin{inparaenum}[\it 1)]
\item Shared Projects,
\item Portable Class Libraries (PCL), and
\item .NET Standard Libraries.\footnote{\url{https://docs.microsoft.com/fr-fr/xamarin/cross-platform/app-fundamentals/code-sharing}}
\end{inparaenum}

Relevant questions ask about these  architectures, specially about the second one. 
There are  questions asking about clarification of those architectures, e.g.,
\qt{What is a Portable Class Library?} (id: 5238955),
\qt{Portable Class Library vs. library project} (id: 28746609), 
or questions about the difference of two architectures, e.g., 
\qt{Xamarin Shared Projects vs Portable class libraries} (id: 23990307).
Other relevant questions focus on problematic of using the PCL architecture, for instance: 
\qt{Unable to resolve assemblies that use Portable Class Libraries} (id: 13871267), or 
\qt{Portable Class library and reflection} (id: 14061291).
Finally, the another group of relevant questions from  this topic relates PCL and concurrency (threads):
\qt{Update UI thread from portable class library} (id: 14427340), 
\qt{Thread.Sleep() in a Portable Class Library} (id: 9251917).

\begin{comment}
\subsection{Conclusions about topic modeling from Xamarin-related questions}
In this section, we have first discovered the main topics discussed on \so and \xf \qa sites.
Then, we have compared those topics with those discovered by \rosen extracted from \so questions about general mobile development.
Our results shows that 27 out of 40 main topics from Xamarin questions are also main topics from question related general mobile app (Android, iOS and Windows Mobile).
Moreover, we found main topics that are excursively related to Xamarin or Microsoft technologies and, consequently,  they were not previously reported by the work of \cite{Rosen2016MDA}.
\end{comment}

\section{Discussion}
\label{sec:discussion}

\subsection{Threats to validity}

\subsubsection{Internal}

\paragraph{Use of tags from posts}
In section \ref{sec:datasets} we applied a technique based on the use of tags for retrieving posts related to Xamarin technology as done by \rosen  \cite{Rosen2016MDA}.
There is a threat when a developer 
\begin{inparaenum}[\it a)]
\item mislabels a post, i.e., tags do not represent the real topic of the posts; or 
\item omits to label it.
\end{inparaenum}
Moreover, we used the titles from posts to capture Xamarin-related posts that were not labelled with Xamarin-related tags. A threat to validity is present when the title does not represent the content of the post. 
To alleviate this threat we manually analyzed a sample of the retrieved posts and  verified the concordance between the title and post's content.

\paragraph{Representation of documents}
We applied the LDA algorithm for modeling topics. 
The input of LDA is a corpus of document (See Section \ref{sec:protocolMainTopics}), where each document contains information of a single post. We decided to, as done by the study we replicated, to only consider  the title of the question due to, according to them: 
\begin{inparaenum}[\it 1)]
\item titles summarize and identify the main concepts being asked in the post, 
\item the body of the question adds non-relevant information rather than the main idea being asked about, and 
\item we are interested in what issues the developers are asking about and adding the answer posts would not make sense
\end{inparaenum} \cite{Rosen2016MDA}.
\paragraph{Configuration of LDA}
LDA needs 4 configuration arguments (alpha, beta, number of topics and number iterations).
Choosing optimal values for those arguments is a difficult task, so, to alleviate this threat, as done by \cite{Hindle2015TMS,Linares-Vasquez2013EAM,Rosen2016MDA}, we tried different configurations to choose, to our best judgment, the best configuration.
The selection criteria used by those works were: 
\begin{inparaenum}[\it a)]
\item  inspection at the average dominant topic probabilities given by the resulting model \cite{Rosen2016MDA}, and 
\item assurance that topics do not have much overlap in top terms, are not copies of each other, and are not share excessive disjoint concepts or ideas \cite{Hindle2015TMS}. 
\end{inparaenum}
In our experiment we decided to use the same configuration that the study we replicated from \rosense.
Reusing the configuration allowed us to compare the topics discovered from Xamarin-related questions against those by \rosen from mobile-related questions.

\paragraph{Label of discovered topics}
As done by the mentioned related works (e.g., \cite{Rosen2016MDA,Linares-Vasquez2013EAM}),  we manually analyzed the results produced by LDA for labeling each topic with a human readable label, based on a set of words from the topic and their probabilistic.
To the best of our knowledge, there is no tool for automatically labeling topics.
To alleviate the threat of mislabeling topics or mismatching of topics from different sources, we have carried out those tasks using a peer-reviewed process and the results are publicly available.

\subsubsection{External}

One potential threat is that sources of information used for studying \qa related to Xamarin technology are not representative. 
To mitigate that threat we selected two sources: \so and \xfse.
The former is one of the most popular and used \qa site by software developers and latter is the official forum of the Xamarin technology.

\subsection{Future Work}

For future work, we plan to continue exploring the two datasets of Xamarin-related questions. 
Future research direction could focus on:

\subsubsection {Combine additional sources of information}
By inspecting the relevant questions of this topic, we notice that many of their accepted answers  include a link to the official Xamarin documentation web site.
For instance, the mentioned most viewed post  (id: 30850510) has an accepted answers that cites a page from the official documentation as source of its response.
That finding triggers some research questions for future work:
\begin{inparaenum}[\it a)]
\item How many posts from \so link to the official documentation in their questions/answers?,
\item How much do (Xamarin) developers ask questions which solution are already included in the documentation?.
\end{inparaenum}
Moreover, we also observe that answers on \so also include links to the Xamarin Forum, the official \qa site of Xamarin.
For example, the question id 24598261 from \so has  an accepted answer based on a post from the Xamarin Forum (id: 66248).\footnote{\url{https://forums.xamarin.com/discussion/comment/66248}}
Other papers have focused on combining different sources of information (e.g., \cite{Zagalsky2016:RCurates, Lee2017,Wang2017Linking,Ye2017DK,Venkatesh2016}).
However, to our knowledge, nobody has linked questions from two \qa sites nor focused on Xamarin source of information.

\subsubsection{Study the architecture of cross-platform mobile apps}
As we discussed on Section \ref{sec:relevantquestions}, relevant questions are about mobile apps architectures and frameworks. 
Future work can study and compare apps developed using different architectures.
For instance, a research could focus on analyzing which is the best option between the architectures 
\emph{Shared Projects} and \emph{Portable class libraries}  in term of, for instance, the expertise  the developers needs to develop and maintain apps with those architectures, or the ease of evolving an application according to the upgrade of the mobile platforms.
Moreover, another possible research for future work could  study the benefices and disadvantages of developing and maintaining cross-platforms apps that used this frameworks (such as Mvvmcross) w.r.t. apps developed without them i.e., using only the Xamarin framework.

\begin{comment}
\subsection{Limitations}

Creation of \xfdb: 
Our fetching engine is not capable of capturing all information the Xamarin forum presents to a registered user.
For instance, a registered user can ``like" an existing answer. 
The Xamarin forum shows the number of like for each answer \emph{only} to registered user. As our fetching engine inspects the pages to fetch as an ``anonymous user" (i.e. it is not logged to the forum site) then the number of likes is not presented.
Moreover, registered user can post images in the comments. The current version of our engine is not capable of retrieving and storing them in our database. 
\end{comment}

\section{Conclusion}
\label{sec:conclusion}

Cross-compiler mobile development frameworks allow mobile developers to create native cross-platform mobile applications with the promise of simplifying the development and maintenance phases by reusing code source across the different platforms.
To study and characterize  
the development and maintenance of cross-platforms apps using cross-compiler frameworks,
in this paper  we present two datasets with questions and answers (Q\&A): one with \qa from the official \xfse, the other with Xamarin-related \qa from \sose.
We then use them in a replication study of \cite{Rosen2016MDA} for discovering the main discussion topics from questions related to Xamarin technology.
We compared the discovered topics against those main topics related to  mobile development discovered by \cite{Rosen2016MDA}. 
We found that a portion of the topics from the Xamarin-related questions were not previously identified when discussing about general mobile applications.
To promote more studies about Xamarin and cross-platform mobile frameworks, the two Xamarin-related \qa datasets are publicly available.

\begin{comment}
For future work, we plan to continue exploring the two datasets of Xamarin-related questions. 
We also plan to study the development process of cross-platform apps using other cross-compiler frameworks such as React Native.
To promote more studies about Xamarin and cross-platform mobile frameworks, our datasets are publicly available.
\end{comment}

\bibliographystyle{plain}
\bibliography{references.bib}
\balance

\end{document}